# From cosmic explosions to terrestrial fires?

Adrian L. Melott[a] and Brian C. Thomas[b]

a Department of Physics and Astronomy, University of Kansas, Lawrence, Kansas 66045 USA. melott@ku.edu

b Department of Physics and Astronomy, Washburn University, Topeka, Kansas 66621, USA. brian.thomas@washburn.edu

ABSTRACT

Multiple lines of evidence point to one or more moderately nearby supernovae, with the strongest signal ~2.6 Ma. We build on previous work to argue for the likelihood of cosmic ray ionization of the atmosphere and electron cascades leading to more frequent lightning, and therefore an increase in nitrate deposition and in wildfires. The potential exists for a large increase in the pre-human nitrate flux onto the surface, which has previously been argued to lead to $CO_2$ drawdown and cooling of the climate. Evidence for increased wildfires exists in an increase in soot and carbon deposits over the relevant period. The wildfires would have contributed to the transition from forest to savanna in northeast Africa, long argued to have been a factor in the evolution of hominin bipedalism.

## 1. Introduction

Astrophysical ionizing radiation events affect the Earth episodically. Effects come from the Sun through flares and solar proton events as well as interstellar supernovae and gamma-ray bursts (reviewed in Melott and Thomas, 2011). Much of the interest regarding terrestrial effects has centered on major mass extinctions (e.g. Melott *et al*., 2004; Melott and Thomas, 2009). Extreme events may eliminate all higher life on Earthlike planets, but are unlikely to cause complete global sterilization (Sloan *et al.* 2017). However, less intense effects are also possible (e.g. Knipp *et al*., 2018), more frequent and therefore more likely to have extant evidence, and still interesting.

There is a great deal of accumulated evidence for supernovae (SN) moderately near the Earth (Knie *et al*., 2004; Fry *et al*., 2016; Mamajek, 2016; Melott, 2016; Wallner *et al*., 2016; Breitschwerdt *et al.*, 2016; Binns *et al*., 2016; Fimiani *et al.*, 2016; Ludwig *et al*., 2016; Erlykin *et al*., 2017). Although the strongest signal is of an event about 2.6 Ma, which coincides within the uncertainties to the Pliocene-Pleistocene (PP) boundary, it may well be a series beginning as long as 7 Ma related to the evacuation of the Local Bubble, a large cavity in the interstellar medium (Breitschwerdt *et al.*, 2016).

The possible terrestrial effect of a nearby SN has long been of interest. More recently, new information on prompt and early emissions from SNe as well as the timing and distance of relatively recent proximate SNe were used as input in a series of computations (Thomas *et al.*, 2016; Melott *et al.*, 2017). Although early work focused on ozone depletion from photon and cosmic ray (CR) irradiation, we have found additional increased importance for the extended arrival of cosmic rays.

We will not repeat here the description of those computations, except for a brief summary. We refer the reader to Melott *et al*. (2019) for a summary or to Melott *et al*. (2017) for more detail.

## 2. Base study methods and conclusions

We assumed a SN of the most common type IIP to have taken place about 2.6 Ma, at a distance of 50 pc. This is taken to be representative, and possibly a member of a series. The Local Bubble had been already formed and the magnetic field is assumed to have been largely expelled and residing within the walls of this structure, which would reflect cosmic rays back into the interior, extending the irradiation time for the Earth.

Cosmic ray propagation in the turbulent field leads to a time-dependent flux at the Earth. We use here the CR flux for case B as computed in Melott *et al.* (2017) at 100y after the arrival of the first photons, and assume that magnetic reflection off the walls will later keep a similar flux inside the cavity of the Local Bubble. The flux of cosmic rays is greatly increased beyond normal levels. For many decades, the primary effect considered for terrestrial life from a nearby SN was ozone depletion (Gehrels *et al*., 2003). We computed this effect based on a series of studies beginning with Thomas *et al.* (2005). In Melott *et al.* (2017), it was estimated that these events would produce about a 50% increase in UVB at the surface, with a range of potential biological impacts (Thomas 2018).

Muons are highly penetrating elementary particles which penetrate up to a km in the ocean, but still by virtue of their large numbers constitute the primary component of irradiation from cosmic rays at the Earth's surface. The large increase in cosmic rays following a nearby SN leads to a similar large increase in muon flux at the Earth's surface (Melott *et al.,* 2017) and in the top km of the oceans (Melott *et al.*, 2019). Our previous work argued that a megafaunal extinction around the PP boundary (Pimiento *et al*., 2017) may have been induced by increased muon flux in the top few hundred meters of ocean (Melott *et al*., 2019).

In this short paper, we discuss additional substantial effects which could also indirectly have had a major effect on the biota, including the evolution of our species. Atmospheric ionization and dissociation from cosmic rays and their subsequent showers of elementary particles led to the ozone depletion mentioned earlier. The ionization may also have had other effects.

### 3. Computation of atmospheric ionization

Atmospheric ionization was computed using tables from Atri *et al.* (2010), originally generated by CORSIKA. CORSIKA (COsmic Ray SImulations for KAscade) is a package combining high- and low-energy interaction models with a specialized scheme for transport in air. The version used for atmospheric ionization was EPOS 1.61, UrQmd 1.3. This gives ionization rate (ions $cm^{-2}$ $s^{-1}$) as a function of altitude for various primary proton energies. The tables in Atri *et al*. (2010) give ionization for primaries with energies from 300 MeV to 1 PeV ($10^{15}$eV) from the ground to 90 km. We used here primary proton energies between 1 GeV and 1 PeV. The ionization rate as a function of altitude was created by convolving the CR flux from Melott *et al.* (2017) with tables from Atri *et al*. (2010).

In Figure 1 we show the energy spectrum of galactic cosmic rays at the present time and at 100 y after the receipt of the first photons from a supernova at 50 pc (Melott *et al*., 2017). Although the computations had open boundary conditions, we expect that the environment would be reflective due to the compression of the magnetic field in the walls of the Local Bubble. Given the size of the structure, long duration of the flux at the 100 y level would be expected. The units are such that equal areas in this plot imply equal amounts of energy deposition at the Earth. Normally, high energy cosmic rays are too low in flux to contribute much to atmospheric ionization. So, besides an overall increase in cosmic ray flux, there is a unique predominance of high energy primaries in the CR flux, up to about a PeV, which are normally not very important for atmospheric ionization. It is important to understand the consequences of this change.

Figure 2 shows results of our analysis of the ionization effects of cosmic ray primaries. We show the ionization flux (ions generated per area per time) per GeV of total incident cosmic-ray energy as a function of altitude and primary energy. It is important to notice that the ionization rate is plotted per GeV unit of primary energy, not per primary. Thus, the expected simple increase in ionization due to more energy per primary is factored out. But, an increase is still seen, particularly in the troposphere. This increase is due to the greater penetration of the high energy cosmic rays in the atmosphere. This will have consequences for activity in the troposphere; our primary interest here is lightning, especially cloud-to-ground lightning. An energy of 1 GeV is representative of the majority of usual galactic cosmic ray primary protons, while the 10

GeV and $10^6$ GeV energies bracket the range most affected by the SN. In the lowest 2 km bin in our model, ionization caused by the higher energy primary protons (the range associated with SN CRs) is about two orders of magnitude higher than that caused by the lower energy associated with galactic CR primaries. There is a great potential for major changes of weather near the surface.

In Figure 3 we show the altitude profile of ionization from galactic CRs under normal conditions and at our fiducial time. (Energetic protons from the Sun mostly do not penetrate down to the troposphere and are not relevant for the issues considered here, as discussed in Usoskin *et al.,* 2011.) The ionization "SNCR" in Figure 3 is computed by convolving the 100 yr, Case B cosmic ray flux as seen in Figure 1 with the ionization as a function of altitude and CR energy as seen in Figure 2. It can be seen that there is a major difference, with ionization in the troposphere about 50 times greater than normal. For comparison we also show our computed ionization profile from the 1956 solar superflare, the hardest and most energetic in the modern era (Usoskin *et al.,* 2011) and therefore amenable to quantitative measurement.

Consequently, the great increase in TeV to PeV cosmic ray flux upon the Earth from an even moderately near supernova should have strong consequences for atmospheric physics near the surface.

## 4. Initiation of cloud-to-ground lightning

It has been proposed that cosmic rays set off electron avalanches in the atmosphere which are the main initiator of lightning (Gurevich *et al.,* 1999, 2008, 2009; Erlykin and Wolfendale, 2010; Chilingarian *et al.*, 2017a; Kumar et al. 2018). Until recently this was an eminently reasonable idea with only circumstantial evidence in its favor. One of the obstacles has been the difficulty of making any *in situ* measurements of the processes initiating lightning. However, Chilingarian *et al*. (2017b) reported observations on Aragats Mountain, Armenia, where the cloud layer is quite low. They found a number of electron avalanches of duration less than a microsecond which were terminated by nearby lightning flashes. This is a smoking gun that makes this a compelling theory, which we will take as our working model.

We note here that it has been proposed that cloud cover is related to cosmic ray flux (Svensmark *et al.,* 2017 and references therein). Since this proposal is highly controversial, we shall only mention it here.

A 50 times increase in atmospheric ionization in the troposphere would clearly make the breakdown and electron cascade much easier, and one could expect a great increase in lightning (e.g. Erlykin and Wolfendale, 2010). Furthermore, the originally isotropic distribution of cosmic rays would result in showers which are much more preferentially vertical, due to the variation in atmospheric column density with angle. So, not only would lightning be enhanced, but cloud-to-ground lightning should be preferentially enhanced. The theory of lightning initiation is not well-developed, and we cannot say that a 50-fold increase in ionization would lead to a 50-fold increase in the number of lightning events. However, the potential is there for a large increase. We note that in the supernova-enhanced regime, there would be of order hundreds of high-energy primaries per $km^2$ per second, so that ionization would be intermediate between continuous and pulsed.

## 5. Effects of increased lightning frequency

One effect of increased atmospheric ionization from supernovae CRs examined . (A.L. Melott, A.L., B.C, Thomas, and B.D. Fields, B.D. (2019b) Climate change via CO2 drawdown from astrophysically initiated atmospheric ionization? arXiv:1810.01995; under review; hereafter Melott *et al.* 2019b) was fixation of nitrogen. Normally, the nitrogen in the atmosphere is largely unavailable to the biota due to the strength of the triple bond in $N_2$. For this reason, nitrogen must be converted into other forms; there is a large literature on the synthesis and deposition of nitrate as a side effect of atmospheric ionization and dissociation (e.g. Thomas *et al*., 2005). Our computations suggested that the increase in available nitrogen would be relatively small, of order 10% from the cosmic ray ionization. However, a large increase in the rate of lighting could provide a strong enhancement in the deposition of nitrate. Lightning is the dominant souce of nitrogen fixation in the pre-industrial world (Schlesinger and Bernhardt 2013, Table 12.3). The computational methods of Thomas et al. (2005) include results from nitrate deposition, mostly in the form of rain or snow. We estimate that it would not be unreasonable for a large increase in the pre-industrial, pre-human nitrogen flux from lightning to result in a similarly large, perhaps several times over nitrate enhancement. This might lead, as discussed in Melott *et al.* (2019b, above) to a $CO_2$ drawdown and cooling of the climate as was observed at the onset of the Pleistocene. We stress that this all depends on there being a large increase in the lightning frequency, which we are unable to estimate.

Lightning is the main initiator of wildfires if one excludes humans (Pausas and Keeley, 2000; Veraverbeke et al. 2017). A large increase in lightning would again be expected to cause a large increase in the number of wildfires started. There has been a coincident worldwide increase in the number of wildfires in since about 7 Ma (Zhou *et al.,* 2014; Bond, 2015), which is probably responsible for the conversion of forest to savanna worldwide (Karp et al. 2018). There has been no good explanation for this increase in fires across various continents and climatic zones (Bond, 2015). We suggest that a global increase in lightning might provide such an explanation.

The evidence for the increase in fires after impact events can be found in soot and other carbon-related sediments (Wolbach *et al*., 1985, 2018). Such evidence might be found here and is in fact cited in Zhou *et al.* (2014) and Bond (2015). While impact-related fires would show a sudden spike, we would expect the SN-CR wildfires to take place over a much more extended period of time, $10^4$-$10^5$ y for each supernova.

The conversion from woodland to savannah has long been held to be a central factor in the evolution of hominins to bipedalism, although more recent thinking tends to view it as a contributing factor (Senut *et al.,* 2018). Thus, it is possible that nearby supernovae played a role in the evolution of humans. This should be borne in mind as more research is done, particularly in the initiation of lightning.

The obvious question to many is the probability of such an event in our future. The nearest apparent Type II supernova progenitor is Betelgeuse, which is likely to go off sometime in the next Myr (Meynet *et al.,* 2013). However, at a distance of ~200 pc, it is not likely to cause serious consequences. We could be affected by a gamma-ray burst from an unseen nearby progenitor with only a few hours' warning (Medvedev and Loeb, 2013), but this is a very low-probability event. The analysis of such events is most useful as a clue to understanding the geological past. For technological humans, solar events are a more reasonable basis for immediate concern (Melott and Thomas, 2012; Knipp *et al.,* 2018).

**Acknowledgments**

We thank Bruce Lieberman and an anonymous referee for helpful suggestions in the presentation.

**Figure Captions**

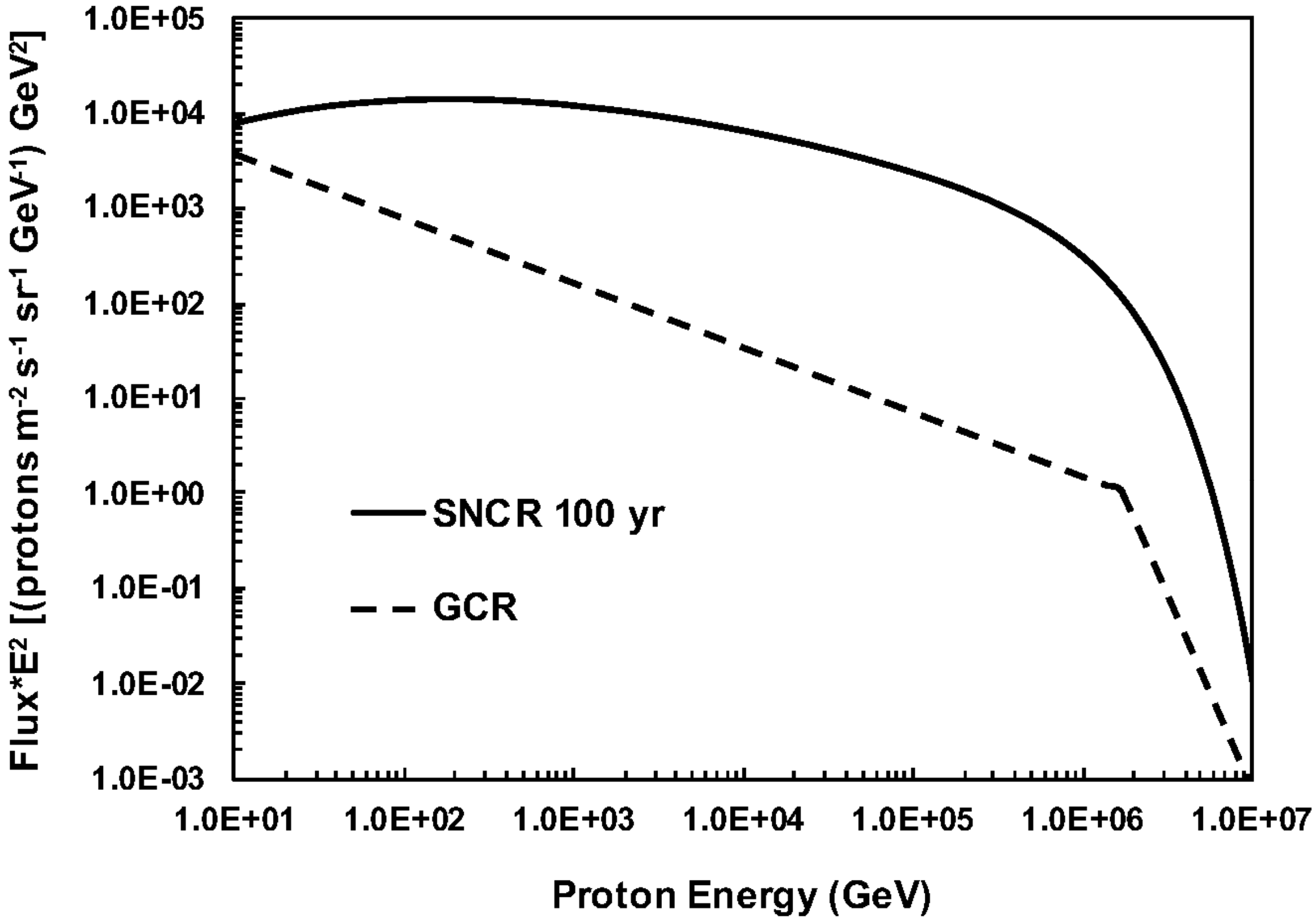

Figure 1 The approximate flux of galactic cosmic rays (dashed line) and the cosmic ray flux computed in Melott et al. (2017) at 100 y after the arrival of the photons from a supernova at 50 pc distance. The units are such that equal areas under the lines correspond to equal total energy flux at the Earth. There is an enhancement of more than two orders of magnitude at the upper range of energies.

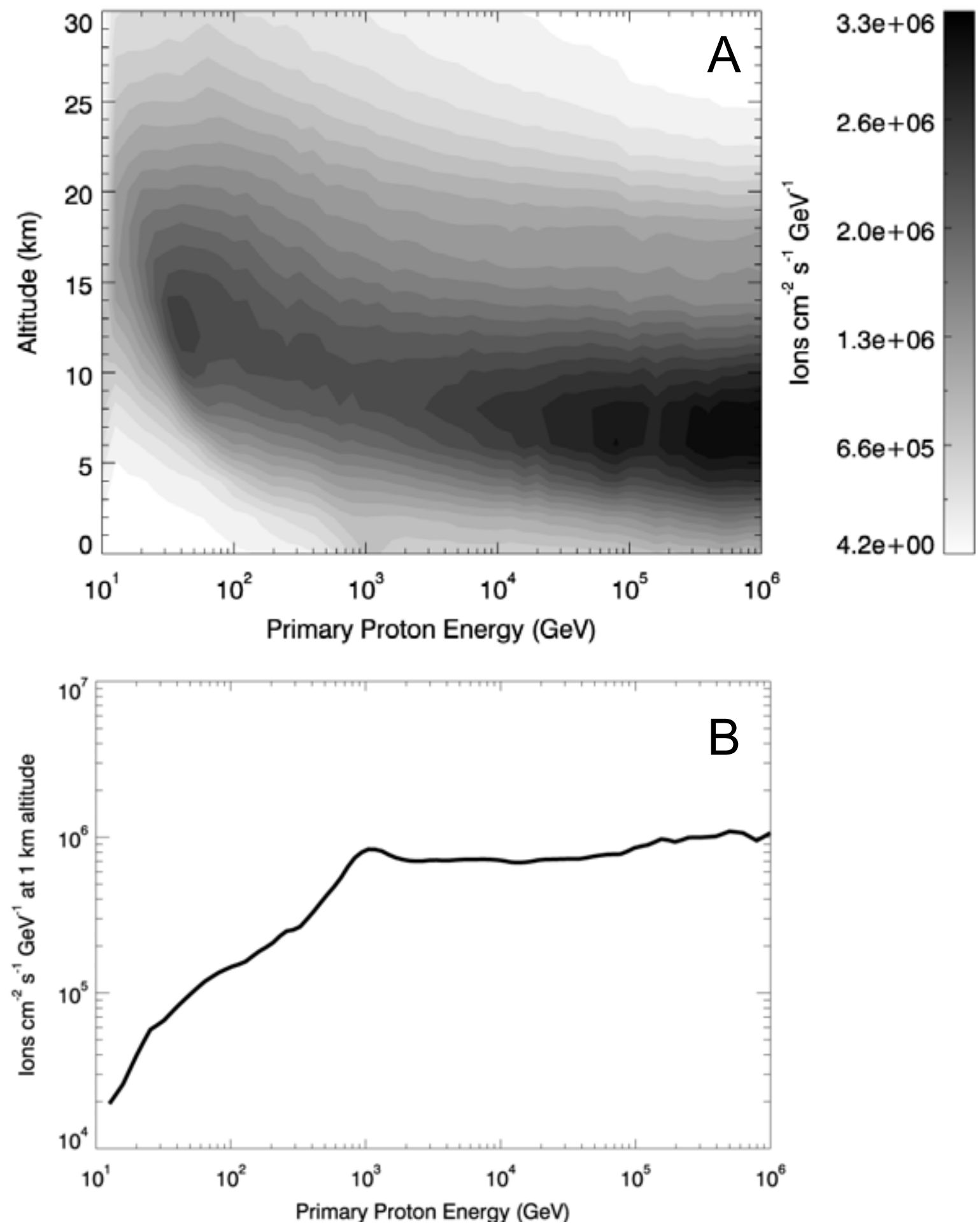

Figure 2: Ionization from the high-energy CRs incident from a moderately nearby supernova ionize the atomosphere at much low altitudes than incident cosmic rays from any other usual source for the Earth. Panel A: ionization rate (flux) per GeV of total incident cosmic-ray energy as a function of altitude and primary energy. We have adjusted to constant energy input units to bring out the altitude effect due to the energy of the primary. Normal cosmic-ray flux peaks near the left boundary of this plot, but moderately nearby supernovae contribute all across it.  Panel B: the ionization rate (flux) in the same units at 1 km altitude, the lowest we resolve. This is most relevant for cloud-to-ground lightning, which should be greatly increased by the additional ionization by high energy cosmic rays.

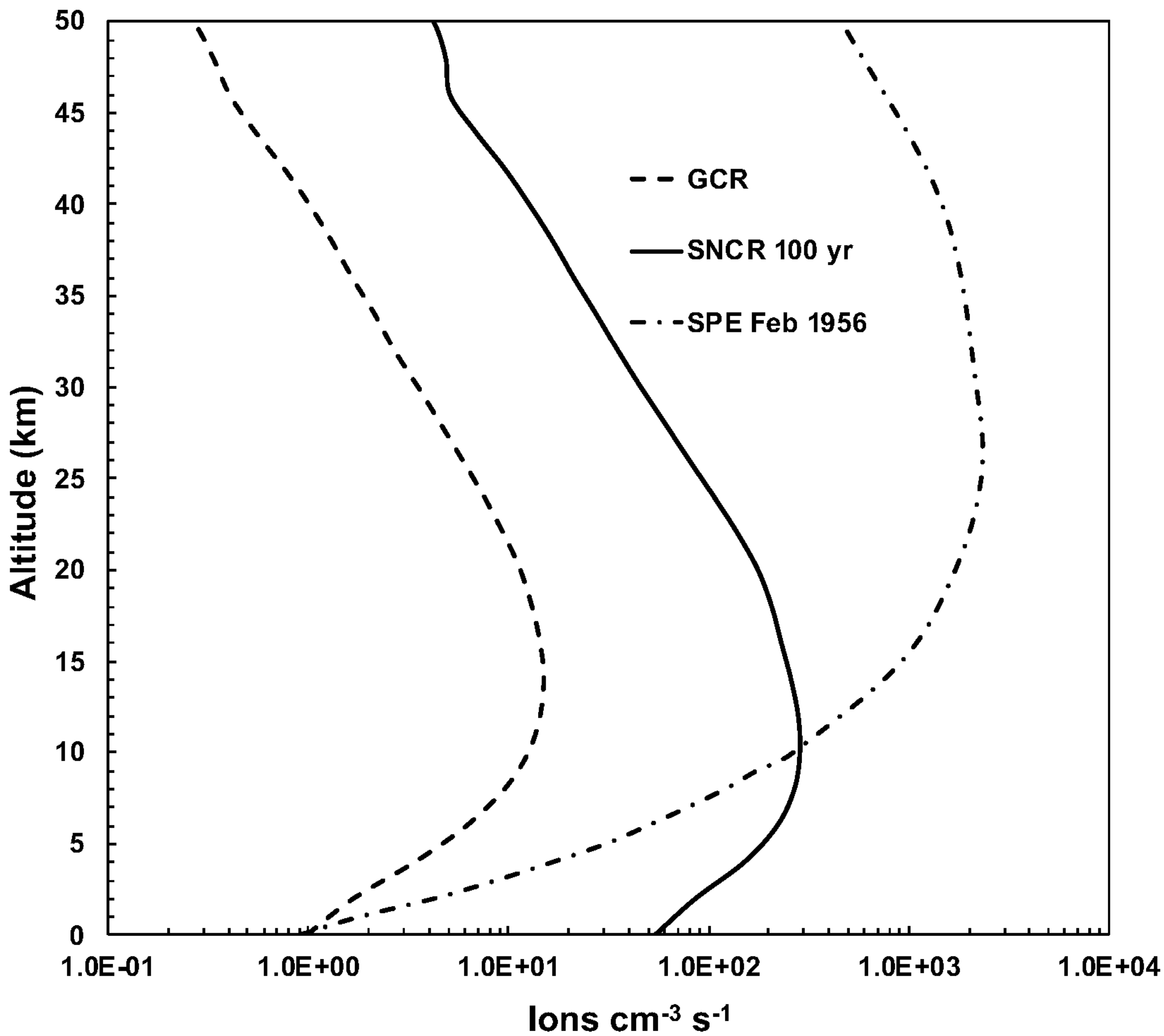

Figure 3: The ionization as a function of altitude for present day typical cosmic rays (dashed line), and the 1956 solar event (dot-dashed line), as compared with the spectrum of incident cosmic rays 100 y after the arrival of photons from a supernova at 50 pc, as computed in Melott *et al.* (2017). It can be seen that ionization in the lowest levels of the atmosphere is increased by a factor of approximately 50.